\documentclass[twocolumn,prl,showpacs,preprintnumbers,amssymb]{revtex4}

\usepackage{epsfig}% Include figure files
\usepackage{dcolumn}% Align table columns on decimal point
\usepackage{bm}% bold math

\def\IC{\bf C}
\def\IZ{\bf Z}
\def\IT{\bf T}
\def\z2z2{$\IC^3/(\IZ_2\times\IZ_2)$}

\def\id{{\bf 1}}

\def\cp{\mbox{\bbbold C}\mbox{\bbbold P}}

\def\a{\alpha}

\def\b{\beta}

\def\d{\delta}\def\D{\Delta}

\def\k{\kappa}
\def\l{\lambda}

\def\p{\pi}

\def\s{\sigma}

\def\th{\theta}

\def\beq{\begin{equation}}\def\eeq{\end{equation}}
\def\beqa{\begin{eqnarray}}\def\eeqa{\end{eqnarray}}
\def\barr{\begin{array}}\def\earr{\end{array}}

\def\wt{\widetilde}

\def\ds {{\del \hspace{-6.4pt} \slash}\;}

\newcommand{\drawsquare}[2]{\hbox{%
\rule{#2pt}{#1pt}\hskip-#2pt%  left vertical
\rule{#1pt}{#2pt}\hskip-#1pt%  lower horizontal
\rule[#1pt]{#1pt}{#2pt}}\rule[#1pt]{#2pt}{#2pt}\hskip-#2pt%  upper horizontal
\rule{#2pt}{#1pt}}% right vertical

\newcommand{\fund}{\raisebox{-.5pt}{\drawsquare{6.5}{0.4}}}%  fund
\newcommand{\Ysymm}{\raisebox{-.5pt}{\drawsquare{6.5}{0.4}}\hskip-0.4pt%
        \raisebox{-.5pt}{\drawsquare{6.5}{0.4}}}%  symmetric second rank
\newcommand{\Yasymm}{\raisebox{-3.5pt}{\drawsquare{6.5}{0.4}}\hskip-6.9pt%
        \raisebox{3pt}{\drawsquare{6.5}{0.4}}}%  antisymmetric second rank
\newcommand{\antifund}{\overline{\fund}}

 \let\br=\bigr

\def\bd{\begin{document}}
\def\ed{\end{document}}
\def\ba{\begin{array}}
\def\ea{\end{array}}
\def\bea{\begin{eqnarray}}
\def\eea{\end{eqnarray}}
\def\ft#1#2{{\textstyle{{\scriptstyle #1}\over {\scriptstyle #2}}}}
\def\fft#1#2{{#1 \over #2}}
\newcommand{\be}{\begin{equation}}
\newcommand{\ee}{\end{equation}}
\newcommand{\eq}[1]{(\ref{#1})}
\def\eqs#1#2{(\ref{#1}-\ref{#2})}
\def\det{{\rm det\,}}
\def\tr{{\rm tr}}
\newcommand{\ho}[1]{$\, ^{#1}$}
\newcommand{\hoch}[1]{$\, ^{#1}$}
\def\ra{\rightarrow}

\def\Xh{\hat{X}}
\def\ah{\hat{a}}
\def\xh{\hat{x}}
\def\yh{\hat{y}}
\def\ph{\hat{p}}
\def\G{{\cal G}}
\def\Dth{{\Delta_\th}}

\def\bk{{\bf k}}
\def\bx{{\bf x}}
\def\br{{\bf r}}
\def\tr{{\rm tr \,}}
\def\Tr{{\rm Tr \,}}
\def\diag{{\rm diag \,}}
\def\tg{{\rm tg \,}}
\def\NPB#1#2#3{Nucl. Phys. B {\bf #1} (19#2) #3}
\def\PLB#1#2#3{Phys. Lett. B {\bf #1} (19#2) #3}
\def\PLBold#1#2#3{Phys. Lett. {#1B} (19#2) #3}
\def\PRD#1#2#3{Phys. Rev. D {\bf #1} (19#2) #3}
\def\PRL#1#2#3{Phys. Rev. Lett. {\bf #1} (19#2) #3}
\def\PRT#1#2#3{Phys. Rep. {\bf #1} C (19#2) #3}
\def\MODA#1#2#3{Mod. Phys. Lett.  {\bf #1} (19#2) #3}
\def\ov{\overline}

\def\preal{{\rm Re\,}}
\def\pim{{\rm Im\,}}

\def\ds{\displaystyle}
\def\yzero{\smash{\hbox{$y\kern-4pt\raise1pt\hbox{${}^\circ$}$}}}
\def\p{\partial}
\def\a{\alpha}
\def\b{\beta}
\def\g{\gamma}
\def\d{\delta}
\def\beq{\begin{equation}}
\def\eeq{\end{equation}}
\def\beqa{\begin{eqnarray}}
\def\eeqa{\end{eqnarray}}
\def\Om{\Omega}
\def\om{\omega}
\def\th{\theta}
\def\vt{\vartheta}
\def\vphi{\varphi}
\def\-{\hphantom{-}}
\def\ov{\overline}
\def\s2{\frac{1}{\sqrt2}}
\def\wh{\widehat}
\def\wt{\widetilde}
\def\oh{\frac{1}{2}}
\def\tr{{\rm tr \,}}
\def\Tr{{\rm Tr \,}}
\def\diag{{\rm diag \,}}
\def\vac{|0 \rangle}
\def\vm{\relax{n_V}}
\def\cc{{\cal C}}
\def\ck{{\cal K}}
\def\ci{{\cal I}}
\def\cu{{\cal U}}
\def\cg{{\cal G}}
\def\cn{{\cal N}}
\def\cam{{\cal M}}
\def\cp{{\cal P}}
\def\ct{{\cal T}}
\def\cv{{\cal V}}
\def\cz{{\cal Z}}
\def\ch{{\cal H}}
\def\cf{{\cal F}}
\def\tv{\tilde v}
\def\Dsl{\,\raise.15ex\hbox{/}\mkern-13.5mu D} %this one can be subscripted
\def\IZ{Z\kern-.4em  Z}
\def\id{{\rm 1}}

\def\ti{\times}
\def\til{\tilde}
\def\eps{\epsilon}
\def\k{\kappa}
\def\A{\Arrowvert}
\def\cw{{\cal W}}
\def\G{\Gamma}
\def\car{{\cal R}}
\def\l{\lambda}
\def\raw{\rightarrow}
\def\Raw{\Rightarrow}
\def\inte{{\bf Z}}
\def\cpx{{\bf C}}
\def\real{{\bf R}}
\def\Lam{\Lambda}
\def\D{\Delta}
\def\cb{{\cal B}}
\def\ca{{\cal A}}

\begin{document}

\preprint{MAD-TH-04-8}
%\twocolumn[\hsize\textwidth\columnwidth\hsize\csname@twocolumnfalse\endcsname

\title{MSSM vacua from Flux Compactifications}

\author{Fernando Marchesano and Gary Shiu}

\affiliation{Department of Physics, 1150 University Avenue, University of Wisconsin, Madison, WI 53706}
\begin{abstract}
We construct the first $D=4$  Minkowski string theory vacua of flux compactification which are (i) chiral, (ii) free of NSNS and RR tadpoles, and (iii) $\cn=1$ or $\cn=0$ supersymmetric. In the latter case SUSY is softly broken by the fluxes, with soft terms being generated in the gauge and chiral sectors of the theory. In addition, the low energy spectrum of the theory is MSSM-like with three generations of chiral
matter, the dilaton/complex structure moduli are stabilized and the supergravity background involves a warped metric.
\end{abstract}
\maketitle

%------%%%%%%%%%%%%%%%%%%%%%%%%%%%%%%%%%%%%%%%%%%%%%%%%%%%%%%%%%%%%%%%%

Moduli stabilization and supersymmetry breaking are perhaps the
two most pressing problems in string phenomenology.
Generic $D=4$, ${\cal N}=1$ string vacua possess a large
number of moduli, which typically remain massless unless supersymmetry is 
broken. Recently, there have been some attempts to
stabilize moduli in string theory by considering compactifications
with fluxes \cite{silver}.
Interestingly,
depending on how
the gauge and chiral sectors are embedded,
these background fluxes can also induce SUSY-breaking
soft terms \cite{susy,susy2}.
While this is an interesting scenario, there is no concrete {\em global}
construction in which the full gauge and matter content of the Standard Model
is embedded in this framework.
In fact, previous attempts \cite{Blumenhagen,cu}
seem to suggest that there is some incompatibility between chirality and
${\cal N}=1$ flux compactifications.

The purpose of this letter is to present the first examples of chiral $D=4$ flux vacua leading to a MSSM-like spectrum. We construct both $\cn = 1$ and $\cn =0$ models, in the latter case SUSY being broken by the presence of the flux. However, both Ramond-Ramond and Neveu-Schwarz tadpoles cancel, so our examples are free of the usual stabilization problems of non-supersymmetric models. Nevertheless, SUSY breaking is felt by the chiral and gauge sectors of the theory, in such a way that soft terms are induced in the effective theory.

The construction of the model is based on type IIB string theory compactified on a ${\bf T}^6/(\inte_2 \times \inte_2)$ orbifold, modded out by an orientifold action. When no RR and NSNS fluxes are turned on, these models are related by T-duality to \cite{bl,csu}. The $\inte_2 \times \inte_2$ generators $\th, \om$ act as $\th: (z_1, z_2, z_3) \mapsto (-z_1, -z_2, z_3)$ and $\om: (z_1, z_2, z_3) \mapsto (z_1, -z_2, -z_3)$. The orientifold modding is given by $\Om \car$, where $\car: (z_1, z_2, z_3) \mapsto (-z_1, -z_2, -z_3)$ and $\Om$ the usual world-sheet parity. The model thus contains 64 O3-planes and 4 O7$_i$-planes, each of them localized at $\inte_2$ fixed points on the $i^{th}$ ${\bf T}^2$ and wrapping the other two.

The above closed string background generates a non-trivial contribution to the Klein bottle string amplitude and hence crosscap tadpoles, which can be canceled by introducing an open string sector. This sector consists of type IIB D$(3+2n)$-branes, filling up $D=4$ Minkowski space and wrapping $2n$-cycles on the compact manifold. The particular example we present below contains D3, D7 and (anti)D9-branes. The latter two may contain a non-trivial magnetic field strength $F=dA$ in 
the internal worldvolume components, as allowed by $D=4$ Poincar\'e invariance. This non-trivial gauge bundle usually reduces the rank of the gauge group and, upon Kaluza-Klein reduction, leads to $D=4$ chiral fermions. This magnetic flux also induces D-brane charge of lower dimension, contributing to the corresponding tadpole. For instance, a D9-brane with magnetic flux will usually have also charges of D7, D5 and D3-brane.

The general framework for constructing magnetized D-brane models in this specific setup was derived in \cite{cu}, whose notation and conventions we will follow. The topological information of a set of $N_a$ D-branes is encoded in six integers $(n_a^i,m_a^i)$: 
$m_a^i$ is the number of times that the D-branes wrap the $i^{th}$ ${\rm T}^2$ and $n_a^i$
is the unit of magnetic flux in that torus. The magnetic field of such D-branes then satisfies
\beq
{m_a^i \over 2\pi} \int_{{\bf T}^2_i} F_a^i = n_a^i.
\label{magnet}
\eeq%%%GS

This notation also describes D7, D5 and D3 branes. For instance, a D7-brane not wrapped on the first torus is expressed as $[(n_a^1,m_a^1)] \times [(n_a^2,m_a^2)] \times [(n_a^3,m_a^3)] = [(1,0)] \times  [(n_a^2,m_a^2)] \times  [(n_a^3,m_a^3)]$, whereas a D3-brane as $ [(1,0)] \times [(1,0)] \times [(1,0)]$. The chiral spectrum arising from two sets of D-branes $a$ and $b$ is determined by the 'intersection product':
\beq
I_{ab} = \prod_{i=1}^{3} \left(n_a^i m_b^i - m_a^i n_b^i \right).
\label{intersection}
\eeq

We also demand the D-brane content to be invariant under the full orientifold symmetry group. In terms of the $\inte_2 \times \inte_2$ symmetry, this implies that D-branes fixed under any of the elements of this group must carry invariant Chan-Paton factors, projecting the initial $U(N_a)$ gauge group to $U(N_a/2)$. On the other hand, invariance under $\Om \car$ implies that to each D-brane $a$ with topological numbers $(n_a^i,m_a^i)$ we must add its image under $\Om \car$, that is, a D-brane $a'$ with magnetic numbers $(n_a^i,-m_a^i)$. This in particular implies that the total D9 and D5-brane charge of the configuration will vanish. Finally, we may consider D-branes fixed by some elements of $\inte_2 \times \inte_2$ and $\Om \car$, which will carry a $USp(N_a)$ gauge group. The general chiral spectrum is summarized in Table \ref{matter}.
\begin{table}[htb] \footnotesize
\renewcommand{\arraystretch}{1.25}
\begin{center}
\begin{tabular}{|c|c|}
\hline
\hspace{1cm} {\bf Sector} \hspace{1cm} &
\hspace{1cm} {\bf Representation} \hspace{1cm} \\
\hline\hline
$aa$   &  $U(N_a/2)$ vector multiplet  \\
       & 3 Adj. chiral multiplets   \\
\hline\hline
$ab+ba$   & $I_{ab}$ $(\fund_a,\antifund_b)$ chirals  \\
\hline\hline
$ab'+b'a$ & $I_{ab'}$ $(\fund_a,\fund_b)$ chirals  \\
\hline\hline
$aa'+a'a$ & $\frac 12 (I_{aa'} - 4 I_{a,O}) \;\;
\Ysymm\;\;$ chirals  \\
          & $\frac 12 (I_{aa'} +  4 I_{a,O}) \;\;
\Yasymm\;\;$ chirals \\
\hline
\end{tabular}
\end{center}
\caption{\small Chiral spectrum on generic magnetized 
D-branes in the $\IT^6/(\inte_2\times \inte_2)$ $\Omega \car$ orientifold. 
$I_{a,O}$ stands for the intersection product between 
$D_a$-brane 
and the orientifold plane.
\label{matter}}
\end{table}           

A consistent string model must satisfy the RR tadpole cancellation conditions, which in this 
particular
setup read:
\beq
\begin{array}{lll} \vspace*{.2cm}
\sum_\a N_\a n_\a^1 n_\a^2 n_\a^3 & = 16, \\\vspace*{.2cm}
\sum_\a N_\a m_\a^1 m_\a^2 n_\a^3 & = -16, \\\vspace*{.2cm}
\sum_\a N_\a m_\a^1 n_\a^2 m_\a^3 & = -16, \\\vspace*{.2cm}
\sum_\a N_\a n_\a^1 m_\a^2 m_\a^3 & = -16.
\end{array}
\label{tadpoles}
\eeq
If we require the D-brane model to be $\cn =1$ supersymmetric we must impose, in addition,
\beq
\sum_i {\rm tan}^{-1} \left( {m_a^i \ca_i \over n_a^i} \right) = 0,
\label{susy}
\eeq
where $\ca_i$ is the area of the $i^{th}$ ${\bf T}^2$ in $\a'$ units. Cancellation of NSNS tadpoles follows from (\ref{tadpoles}) and (\ref{susy}).

In Table \ref{Ymodel} we present an example of a magnetized D-brane model satisfying all the previous requirements, while still yielding a semi-realistic chiral spectrum. Indeed, is easy to see that RR tadpoles (\ref{tadpoles}) are satisfied by simply imposing $g^2 + N_f = 14$. Notice that this give us an upper bound $g \leq 3$. Interestingly enough, $g$ will turn out to be the number of families in our model. On the other hand, the supersymmetry conditions (\ref{susy}) can be satisfied by simply choosing
\beq
\begin{array}{c}\vspace*{.15cm}
\ca_2 = \ca_3 = \ca\\
{\rm tan}^{-1} (\ca/3) + {\rm tan}^{-1} (\ca/4) + {\rm tan}^{-1}(\ca_1/2) = \pi
\end{array}
\eeq
\begin{table}[htb]
\renewcommand{\arraystretch}{1.25}
\begin{center}
\begin{tabular}{|c||c|c|c|}
\hline
 $N_\a$  &  $(n_\a^{1},m_\a^{1})$  &  $(n_\a^{2},m_\a^{2})$   
&  $(n_\a^{3},m_\a^{3})$ \\ 
\hline\hline $N_a = 6+2$ & $(1,0)$ & $(g,1)$ & $(g,-1)$  \\
\hline $N_b=2$ & $(0,1)$ &  $ (1,0)$  & $(0,-1)$ \\
\hline $N_c=2$ & $(0,1)$ &  $(0,-1)$  & $(1,0)$  \\
\hline \hline
$N_{h_1}= 2$ & $(-2,-1)$  &  $(3,1)$ & $(4,1)$  \\
\hline $N_{h_2}= 2$ & $(-2,-1)$ & $(4,1)$ & $(3,1)$ \\
\hline $8N_{f} $ & $(1,0)$ &  $(1,0)$  & $(1,0)$  \\
\hline \end{tabular}
\caption{\small D-brane magnetic numbers giving rise to an $\cn=1$ chiral spectrum. For the particular value of $g=3$, we recover the MSSM spectrum as a subsector of the theory. \label{Ymodel}}
\end{center}
\end{table}
The chiral spectrum of this model can be computed from Table \ref{matter} and is easy to show that it leads to a $SU(4) \times USp(2) \times USp(2) \times U(1)^3 \times USp(8N_f)$ gauge group. Not all of these Abelian factors will remain as gauge symmetries of the low energy theory, since some of them will couple to closed string RR fields mediating a generalized Green-Schwarz (GS) mechanism which cancels mixed $U(1)_\a-SU(N_\b)^2$ and gravitational anomalies. The corresponding gauge boson will acquire a Stueckelberg mass and the previous gauge symmetry will remain as a global symmetry of the effective Lagrangian. In addition, the $USp(8N_f)$ factor will only remain as such when the $8N_f$ D3-branes are placed on top of an orientifold singularity. By moving them away, we can Higgs this group down to $U(1)^{2N_f}$. After all these considerations we recover a gauge group of the form
\beq
SU(4) \times SU(2) \times SU(2) \times U(1)'  \times U(1)^{2N_f},
\label{gauge}
\eeq
where we have made use of the identity $USp(2) \simeq SU(2)$. The Abelian factor not coming from D3-branes and surviving the GS mechanism is given by $U(1)' = U(1)_a - 2g\, [U(1)_{h_1} - U(1)_{h_2}]$. We thus find a Pati-Salam gauge group plus some additional Abelian factors.

The chiral spectrum of the theory contains $g$ generations of quarks and leptons arranged in Pati-Salam multiplets, as well as a MSSM Higgs sector. In addition, there is some exotic chiral matter charged both under the Pati-Salam gauge group and the $U(1)'$ Abelian factor. This exotic spectrum greatly simplifies for the particular case $g=3$, which is displayed in Table \ref{Yspectrum}.
\begin{table}[htb]
\renewcommand{\arraystretch}{1.25}
\begin{center}
\begin{tabular}{|c|c|c|c|}
\hline Sector &
 Matter  & $SU(4) \ti SU(2) \ti SU(2)$  &  $U(1)'$   \\
\hline\hline (ab) & $F_L$ &  $3(4,2,1)$ & 1    \\
\hline (ac) & $F_R$   &  $3( {\bar 4},1,2)$ & -1  \\
\hline (bc) & $H$    &  $(1,2,2)$ &  0    \\
\hline
\hline ($ah_1'$) &    &  $6(\bar{4},1,1)$ &  $5$  \\
\hline ($ah_2$) &    &  $6({4},1,1)$ &  $-5$  \\
\hline ($bh_1$) &    &  $8(1,2,1)$ &  $6$  \\
\hline ($bh_2$) &    &  $6(1,2,1)$ &  $-6$  \\
\hline ($ch_1$) &    &  $6(1,1,2)$ &  $6$  \\
\hline ($ch_2$) &    &  $8(1,1,2)$ &  $-6$  \\
\hline 
\end{tabular}
\caption{\small Three generation Pati-Salam $\cn=1$ spectrum derived from the D-brane content of table \ref{Ymodel}. We display the chiral exotics charged under the non-Abelian factors. \label{Yspectrum}}
\end{center}
\end{table}

In addition to the chiral multiplets presented in Table \ref{Yspectrum}, there are matter fields not charged under the Pati-Salam gauge group. Of particular interest are those in the $(h_1h_2')$ sector of the theory, of multiplicity 196 and uncharged under the gauge group (\ref{gauge}). These chiral multiplets parametrize a moduli space of flat directions in the $\cn=1$ effective theory, which can acquire a non-vanishing v.e.v. From the D-brane physics perspective, this amounts to (anti)D9-brane recombination $h_1 + h_2'  \raw h$ which only breaks the $U(1)_{h_1} + U(1)_{h_2}$ gauge factor, already massive at low energies. Thus, performing this Higgsing does not affect the low energy gauge group, and in particular the Pati-Salam sector. It does, however, have an important effect on the chiral spectrum of the theory. Indeed, the recombined D-brane system $h$ is nothing but an anti-D9-brane with a gauge bundle which is a deformation of the direct sum of the bundles $h_1$ and $h_2$.  As such, it contains a $U(1)$ gauge theory and posses the magnetic charges $[{\bf Q}_h] = [{\bf Q}_{h_1}] +  \Om \car [{\bf Q}_{h_2}]$. We can compute the chiral spectrum of this deformation of Table \ref{Ymodel} with the same topological formulae of Table \ref{matter}, finding that the final theory has the extremely simple chiral content of Table \ref{Yspectrum2}.
\begin{table}[htb]
\renewcommand{\arraystretch}{1.25}
\begin{center}
\begin{tabular}{|c|c|c|c|}
\hline Sector &
 Matter  & $SU(4) \ti SU(2) \ti SU(2)$  &  $U(1)'$  \\
\hline\hline (ab) & $F_L$ &  $3(4,2,1)$ & 1    \\
\hline (ac) & $F_R$   &  $3( {\bar 4},1,2)$ & -1  \\
\hline (bc) & $H$    &  $(1,2,2)$ &  0    \\
\hline
\hline ($bh$) &    &  $2(1,2,1)$ &  $6$  \\
\hline ($ch$) &    &  $2(1,1,2)$ &  $-6$  \\
\hline 
\end{tabular}
\caption{\small $\cn=1$ spectrum derived from the D-brane content of Table \ref{Ymodel} after D-brane recombination. There is no chiral matter arising from $ah$, $ah'$, $hh'$ or charged under $U(1)^{2N_f}$. \label{Yspectrum2}}
\end{center}
\end{table}

This spectrum can be further simplified by performing additional Higgsing, but we will not pursue this direction here. Notice that the Pati-Salam sector of the theory (i.e., the upper part of the Table \ref{Ymodel}) is not involved in the D-brane recombination process, and hence remains as a simple sector of three sets of D7$_1$, D7$_2$ and D7$_3$-branes, the first of them with a (factorisable) gauge bundle in their internal worldvolume. For the particular choice $g=3$, this sector is nothing but a T-dual version of the local intersecting D-brane model constructed in \cite{yukis1}, which captures most of the semi-realistic physics features of the present construction. As shown in the local construction in \cite{yukis1}, the {\em Pati-Salam spectrum can be broken to a $SU(3) \times SU(2) \times U(1)_Y$ MSSM spectrum} by simple Higgsing, with the usual Higgs sector and a $\mu$-term. Moreover, the simplicity of this local model allows us to compute the MSSM Yukawa couplings analytically in terms of theta functions \cite{yukis1,yukis2}, with the result of one heavy generation of quarks and leptons.

Although this simple construction contains a low energy spectrum remarkably close to the MSSM, it still lacks some basic ingredients for constructing a fully realistic model of particle physics. 
First, the theory contains a large number of closed string moduli fields, which translate into unobserved massless fundamental scalars in the effective theory description. Moreover, supersymmetry must be broken at some level, in order to generate a non-vanishing masses for Standard Model superpartners. Notice, as well, that the hidden sector of this theory does only contain Abelian factors, and hence the usual field theory mechanism for SUSY breaking via gaugino condensation in a strongly-coupled hidden sector would in principle not work.

It has been recently realized, however, that the two generic problems of string constructions mentioned above can be simultaneously solved by introducing non-trivial RR and NSNS 3-form fluxes in the theory. These type IIB fluxes indeed generate a scalar potential for the dilaton and complex structure moduli fields, freezing them at some particular values \cite{GVW,drs,gkp}. In addition, supersymmetry can be broken by some particular components of the flux, while the cosmological constant remains zero (to lowest order) due to the no-scale structure of the potential \cite{gkp}.
Nevertheless, part of the open string sector, in particular D7-branes and chiral matter between them, feels such flux supersymmetry breaking, which generates soft SUSY breaking in the low energy effective Lagrangian \cite{susy2}. This soft terms structure, moreover, turns out to be particularly simple and have suggested interesting solutions to several problems associated with general soft SUSY breaking patterns \cite{fluxed}.

An important feature of the model constructed above
is that it not only allows to embed a semi-realistic gauge and chiral sector, but also can include the presence of nontrivial RR ($F_3$) and NSNS ($H_3$) 3-form fluxes. Indeed, an antisymmetric field background of the form $G_3 = F_3 - \tau H_3$, $\tau = a + i/g_s$ being the type IIB axion-dilaton coupling, carries a D3-brane RR charge given by
\beq
N_{\rm flux} =  {i \over (4\pi^2 \a^\prime)^2} \int_{\cam_6} {G_3 \wedge \overline{G}_3 \over 2\pim \tau} \in 64 \inte
\label{RRcharge}
\eeq
the multiplicity of 64 due to the quantization conditions of $F_3$ and $H_3$ in this orientifold background \cite{cu}. A 3-form flux satisfying the BPS-like ISD condition ($*_6 G_3 = i G_3$) will carry both RR and NSNS positive D3-brane charges. As a result, the D3-brane tadpole conditions for the previous model get modified to $g^2 + N_f + \frac{1}{16}N_{\rm flux} = 14$. This allows for several solutions of the form
\begin{itemize}

\item $g = 3$, $N_{\rm flux}= 64$, $N_f = 1$

\item $g = 2$, $N_{\rm flux}= 2 \cdot 64$, $N_f = 2$

\item $g = 1$, $N_{\rm flux}= 3 \cdot 64$, $N_f = 1$

\end{itemize}

As an example, let us consider the last solution. It can be achieved by considering the 3-form flux
\beq
G_3\, =\, \frac{8}{\sqrt{3}}\, e^{-\frac{\pi i}{6}}\, ( d{\ov 
z}_1dz_2dz_3 +
dz_1d{\ov z}_2dz_3 + dz_1dz_2d{\ov z}_3 )
\label{susyflux}
\eeq
which fixed the untwisted complex structure moduli and the dilaton to $\tau_1 = \tau_2 = \tau_3 = \tau = e^{2\pi i/3}$. The flux (\ref{susyflux}) is a combination of $(2,1)$ 3-forms, and hence the closed string background preserves $\cn=1$ supersymmetry \cite{granapol}. We thus conclude that, unlike what previous attempts may have suggested \cite{Blumenhagen,cu}, it is indeed possible to find chiral $\cn=1$ string theory vacua involving 3-form fluxes and magnetized D-branes. 
A more realistic example is given by choosing $g=3$ and $N_{\rm flux} = 64$. The latter can be achieved by choosing 
\vspace*{.1cm}
\begin{center}
$G_3\, =\, 2\, ( d{\ov z}_1dz_2dz_3 + dz_1d{\ov z}_2dz_3 + dz_1dz_2d{\ov z}_3 + d{\ov z}_1 d{\ov z}_2 d{\ov z}_3)$
\end{center}
\vspace*{.1cm}
with dilaton and complex structure moduli $\tau_1 = \tau_2 = \tau_3 = \tau = i$ \footnote{This choice of complex dilaton gives $g_s = 1$ and string perturbation theory may seem no longer reliable. However, there are also solutions with arbitrarily small $g_s$.}. Since the above $G_3$ flux contains a $(0,3)$ component, supersymmetry will be broken by this closed string background field. Although NSNS tadpoles are still canceled, this component of the flux will generate a non-vanishing gravitino mass as well as soft terms for the MSSM effective Lagrangian \cite{susy2}.

%Finally, although 
Although we have performed this  construction on the particular $\inte_2 \times \inte_2$ background of \cite{cu}, similar techniques
%it is easy to see that it can also be 
can be carried out in the alternative $\inte_2 \times \inte_2$ orientifold considered in \cite{Blumenhagen}, and other compactifications. Hence, the construction of MSSM flux vacua does not seem to be limited to a particular closed string background. We will present these 
%and other 
flux compactification examples 
%elsewhere
in  \cite{MS}.
 
Having constructed these $\cn=1$ and $\cn=0$ compact models, both being
chiral examples of the warped metric solutions found in \cite{gkp}, one may naturally wonder what is their lift to F-theory. One can also consider heterotic \cite{drs} and type IIA \cite{kahler} duals of the above constructions, both probably involving non-K\"ahler geometries.

These explicit models, in particular those with $\cn=0$, also provide a good starting point for studying the phenomenological possibilities of $D=4$ chiral flux compactifications. In particular, how the general pattern of SUSY-breaking soft terms deduced from \cite{susy2} apply to these particular compact examples. Although a naive general analysis suggest a typical scale of order $\a'/\sqrt{Vol(T^6)}$ for these terms which favors an intermediate string scale $M_s = 10^{11} GeV$, a non-trivial, inhomogeneous warp factor may change this situation. It would be interesting to see if this is the case. Finally notice that, besides realizing flux-induced SUSY breaking, these models contain basic ingredients of the recent proposal for constructing de Sitter vacua from string theory \cite{KKLT}, so it would be nice to find examples combining both scenarios. We hope to report on these issues in the future.

To sum up, we have constructed some examples of string vacua that gather several essential ingredients for building a fully realistic string theory model.  We consider both D-branes and RR and NSNS 3-form fluxes in the same construction. The D-brane sector introduces a non-Abelian gauge group and chiral matter charged under it, allowing for MSSM-like spectra. The 3-form flux background generates a potential for the dilaton and complex structure moduli of the compactification, freezing them to some particular value. If in addition we consider a supersymmetry breaking flux, soft terms would be induced for the gauge and chiral sectors of the theory. We find it
quite remarkable that all these interesting features can be realized in the same string theory construction, in the absence of NSNS tadpoles, and with such a partial knowledge of the whole set of string/M-theory vacua.

\acknowledgments

%\vspace*{.2cm}

%\centerline{\bf Acknowledgements}

%\vspace*{.1cm}

\vspace*{.05cm}

We wish to thank L.~Ib\'a\~nez and A.M.~Uranga for useful discussions.
This work was supported in part by NSF CAREER Award No.~PHY-0348093, and a Research Innovation Award from Research Corporation.

\end{document}